\documentclass[11pt,psfig]{article}
\usepackage{amssymb}
\usepackage{amsmath,amsthm}
\usepackage{amsfonts}
\usepackage{graphicx}
\usepackage{graphics}
\numberwithin{equation}{section}
\newtheorem{Theorem}{Theorem}[section]

\usepackage{hyperref}

\title{ Smooth Transition  HYGARCH Model: Stability and Forecasting}
\author{ Ferdous Mohammadi$^*$  and  Saeid Rezakhah\footnote{
Department of Statistics, Faculty of Mathematics and Computer Science, Amirkabir University of
Technology,  424 Hafez Avenue, Tehran 15914, Iran}$\,\;$\footnote{Corresponding author   $\;\;$
 email: rezakhah@aut.ac.ir
}}
\vspace{0.5cm}
\begin{document}
\maketitle

\begin{abstract}
 HYGARCH process is the commonly used long memory process in modeling the long-rang dependence in volatility.
 Financial time series are characterized by transition between phases of different volatility levels. The smooth transition HYGARCH (ST-HYGARCH) model is proposed to model time-varying structure with long memory property. The asymptotic behavior of the second moment is studied and an upper bound for it is derived. A score test is developed to check the smooth transition property. The asymptotic behavior  of the  proposed model and the score test is examined by simulation. The proposed model is applied  to the  \textit{S}\&\textit{P}500 indices for some period which  show evidence of smooth transition property and demonstrates out-performance of the ST-HYGARCH  than HYGARCH in forecasting.
\end{abstract}
%


\section{Introduction}
In the past four decade, modeling and forecasting the time-varying conditional variance or volatility of the financial time series has received vast attentions. In the financial time series the periods of large volatility followed by periods of low volatility. This characteristic led to the idea that volatility is predictable. The ARCH and GARCH model introduced respectively  by Engle \cite{engle 1982} and Bollerslev \cite{bollerslev 1986} are quite successful in modelling the dynamic volatility of the financial time series. It has been shown that the volatility of the financial time series tend to display long memory and theirs correlations stay positive for long lags and decay slowly to zero ( see Green and Fielits \cite{green and fielits 1997}, Ding et al.\cite{ding 1993}, Kokoszka and Taqqu \cite{kokoszak and taqqu 1996} and cont \cite{cont 2001}). In the other hand the autocorrelation function (ACF) of the GARCH model decays exponentially which implys short memory and can not capture the long memory in volatility. Baillie et al.\cite{baillie 1996} proposed the FIGARCH model to overcome this shortcoming. FIGARCH process exhibits the hyperbolic decaying of the ACF. However the variance of the FIGARCH model always dose not exist. Davidson \cite{davidson 2004} extended the FIGARCH model and proposed the HYGARCH model. The conditional variance of the HYGARCH model is a convex combination of those of  the GARCH model and the FIGARCH model respectively with weights 1-w and w. HYGARCH process has the desired property of the finite variance  as the GARCH process  while at the same time its ACF decays hyperbolically. It  is  successful in modeling long memory dynamic of the volatility (Davidson \cite{davidson 2004}, Tang and Shich \cite{tang and shieh 2006} and Niguez and Rubia \cite{niguez 2006} ). 
Conrad \cite{conrad 2010} derived the conditions for the non-negativity in conditional variance of the HYGARCH model. Li et al.\cite{li 2011} proposed a simplified score test for exponential decay against hyperbolic decay in the HYGARCH process. Kwan et al.\cite{kwan 2011} proposed a threshold HYGARCH model to jointly capture the long memory and regime switching between phases of low volatility and phases of high volatility. Li et al.\cite{li 2013} proposed a new hyperbolic model where it has to mixture components with bernoulli coefficients. Li et al.\cite{li 2015} 
proposed a hyperbolic model that has a form nearly the FIGARCH process while allowing the existence of finite variance.
Empirical evidences show that  economic or political events may cause the structure of the volatility changes over time. This means that a stationary model for volatility may not be adequate. 
Models which allow for state-dependent or regime-switching behavior have been most appropriate and popular in application to financial time series. Smooth transition (ST) models are of the regime-switching models. This class of models deal with the structural changes in volatility and assume the smoothly non-stationary process. 
For review of the ST model refer to Granger and Ter\"asvirta \cite{granger and terasvirta 1993}, Ter\"asvirta \cite{terasvirta 1998}, Gonzales-Rivera \cite{gonzales 1998} and Lubrano \cite{lubrano 2001}. 
 The ST models are an extension of the two-regime models because they allow intermediate regimes. The main advantages of ST models are that, it is not require to determine the number of the regimes (states) a priori, where it is a challenging problem and may led to over or under-fitting the number of the regimes. The ST model changes smoothly according to the transition variable rather than jump suddenly between discrete states as  markow switching models.
 
  To impose the smooth transition structure for the conditional variance of HYGARCH model we allow the weights of  the convex combination to be time dependent  and logistic function of past observations. We denote this model by ST-HYGARCH. The continuity property of the logistic function which relies on the different transition variables  can led to different degrees of the smoothness.  The ST-HYGARCH model allows to conduct the smooth transition between the influence of  extreme regimes and long memory feature of the volatilities in parsimonious way. We follow the method of  Alemohammad and Rezakhah \cite{alemohammad 2016} to derive the necessary and sufficient condition for the second moment to be asymptotically bounded. We develop a score test also called Lagrange multiplier test  to check  the presence of the smooth transition property in the model. By simulation  the size and power of the proposed test are evaluated. The parameters of the model are estimated via maximum likelihood  method. Real data of the \textit{S}\&\textit{P}500 indices  for some special period  which approving evidence of the smooth transition property  are considered which show that  the ST-HYGARCH  out-performs the HYGARCH in forecasting.
   
   The structure of the paper is as follows. Section 2 present the ST-HYGARH model. In section 3 the stability of the model is analyzed. The maximum likelihood estimators are calculated in section 4. In section 5 the score test for investigation the smooth transition property are discussed. Section 6 is devoted to the simulation. Real example is considered in section 7. Finally section 8 concludes the paper.

\section{The Model}
The first order HYGARCH model is  defined as: 
$$y_t=\epsilon_t \sqrt{h_t} $$
\begin{equation}
h_t=\dfrac{\gamma}{1-\beta}+\{1-\dfrac{1-\delta B}{1-\beta B}[1-w+w(1-B)^d]\}y_t^2\label{1},
\end{equation}
where $\{\epsilon_t\}$ are an \textit{iid} random variables with mean 0 and variance 1.  $\gamma>0$, $\beta\geq0$,  $\delta\geq0$. B is the back-shift operator. $0\leq d \leq1$ and $w\geq0$ are the memory and weight parameters respectively. The hyperbolic memory of the model originated from the structure $(1-B)^d=1-\Sigma_{i=1}^{\infty}\pi_i B^i$ where $\pi_i=\dfrac{d\Gamma(i-d)}{\Gamma(1-d)\Gamma(i+1)}$ for $0<d<1$. $\sqrt{h_t}$ is the volatility of the process and $Var(y_t|\Upsilon_{t-1})=h_t$ where $\Upsilon_{t-1}$ be the information set available at time t-1. It is assumed that $\epsilon_t$ independent from $\Upsilon_{t-1}$. 

(\ref{1}) can be rewritten as:
$$ h_t=(1-w)h_{1,t}+wh_{2,t}$$ 
with 
\begin{equation}
h_{1,t}=\gamma+\beta h_{1,t-1}+(\delta-\beta)y^2_{t-1} \label{2}
\end{equation}
and
\begin{equation}
h_{2,t}=\gamma+\beta h_{2,t-1}+[1-\beta B-(1-\delta B)(1-B)^d]y^2_{t}.\label{3}
\end{equation}
where  (\ref{2}) and (\ref{3}) are the conditional variances of the  GARCH(1,1) and FIGARCH(1,d,1) respectively. 
$w$ which determine the impression of the GARCH and FIGARCH conditional variances in the HYGARCH model is constant over time and this dose not conform with the time-varying structure of the volatilities specifically in financial time series. Hence we improve the dynamic behavior of the HYGARCH model by introducing time-varying weights.

\subsection{The Smooth Transition HYGARCH Model}
A time series $\{y_t\}$ follows a first order ST-HYGARCH model, if
\begin{equation}
 y_t=\sqrt{h_t} \epsilon_t  \label{4}
\end{equation}
 
\begin{equation}
h_{t}=(1-w_{t})h_{1,t}+w_th_{2,t} \label{5}
\end{equation}
 where
 \begin{equation}
h_{1,t}=a_{0}+a_{1}h_{1,t-1}+a_{2}y_{t-1}^2 ,\label{6}
\end{equation}

\begin{equation}
h_{2,t}=b_{0}+b_{1}h_{2,t-1}+[1-b_{1}B-(1-b_{2}B)(1-B)^{d}]y_{t}^2  \label{7}
\end{equation}
 and
 \begin{eqnarray}
 w_t=\dfrac{exp(-\gamma z_t)}{1+exp(-\gamma z_t)} \label{8}
 \end{eqnarray}
 where$ \{\epsilon_t \}$ are iid standard normal variables. We impose the constraints $a_0, a_1, a_2, b_0 >0$ and $0<b_2\leq b_1 \leq d<1$ (Chung \cite{chung}). These conditions are sufficient to guarantee strictly positive conditional variance. $w_t$ is a logistic function that is monotonically increasing function and bounded between 0 and 1. $\gamma$ is called the smoothness or slope parameter and determines the speed of  transition between different regimes. It commonly assumed positive. When $\gamma\rightarrow\infty$ the logistic function become a step function and the ST-HYGARCH model falls in the class of the threshold model. $z_t$ is known as the transition variable. There are several possible choises for $z_t$. For example if $z_t=y_{t-k}$ (for a suitable $k$) then the differences in the dynamic of the conditional variance are modelled according to the size and sign of the past shock, or  $z_t=h_{t-k}$ means that regime switching down according to the past volatility. $z_t$ can be a nonlinear function of the previous observations. It can also be an exogenous variable. In financial literature  several choices for $z_t$ is proposed, for example an international market return, economic index or the past cumulated returns (see Dijk et al.\cite{dijk 2002}, Grelach and Cheen \cite{grelach and cheen 2008} and McAleer et al.\cite{mcaleer 2008}).
 The extreme regimes occur when $w_t\rightarrow1$ as $z_t\rightarrow\infty$ (wherein the ST-HYGARCH model tend to FIGARCH model) and $w_t\rightarrow0$ as $z_t\rightarrow-\infty$ (wherein ST-HYGARCH model tend to GARCH model). ST-HYGARCH model is a member of the regime switching models class that allows the time series to move between extreme regimes where transition is smooth and governed by $z_t$. The regime that  occurs at time $t$ determined by $z_t$ and the associated value of $w_t$ so ST-HYGARCH model is capable to generate changes in the dynamic behavior of the volatilities.

 \section{Stability} 
 One of the main property for any new proposed model is the stability of the model. Here stability refers to the behavior of the second moment of model. In this section we show that under some conditions the second moment of the ST-HYGARCH model is asymptotically bounded. The second moment of the model  calculated as:
 \begin{equation}
 E (y_t^2)=E(h_t\epsilon_t ^2)=E(h_t) \label{9}
 \end{equation}
 Note that  relation (\ref{7}) can be rewritten as:
 
\begin{equation}
h_{2,t}=b_{0}+b_{1}h_{2,t-1}+(b_2-b_1+\pi_1)y_{t-1}^2+\sum_{i=0}^{\infty}(\pi_{i+2}-b_2\pi_{i+1})B^iy_{t-2}^2.\label{10}
\end{equation}
 So using (\ref{10}) we have
 \begin{align}
 E(h_t)&=E((1-w_t)h_{1,t}+w_th_{2,t}) \nonumber \\ 
 &=a_0+\underbrace{(b_0-a_0)E(w_t)}_{I}+a_1\underbrace{E((1-w_t)h_{1,t-1})}_{II}+b_1\underbrace{E(w_th_{2,t-1})}_{III}+ \nonumber\\
 &=\underbrace{(b_2-b_1+\pi _1-a_2)E(w_ty_{t-1}^2)}_{IV}+a_2E(y_{t-1}^2)+\sum_{i=0}^{\infty}(\pi_{i+2}-b_2\pi_{i+1})\underbrace{E(w_ty_{t-2-i}^2)}_V \label{11}
 \end{align}

since $0<w_t<1$, hence an upper bounds for I, II, III,IV and V are obtained as:
\begin{align}
&(b_0-a_0)E(w_t)\leq |b_0-a_0| \nonumber \\
&E((1-w_t)h_{1,t-1})\leq E(h_{1,t-1}) \nonumber \\
&E(w_th_{2,t-1})\leq E(h_{2,t-1}) \nonumber \\
&(b_2-b_1+\pi _1-a_2)E(w_ty_{t-1}^2)\leq |b_2-b_1+\pi _1-a_2|E(y_{t-1}^2)  \nonumber \\
&E(w_ty_{t-2-i}^2)\leq E(y_{t-2-i}^2).   \label{110} 
\end{align} 
By replacing the obtained upper bounds (\ref{110}) in (\ref{11}) an upper bound for $E(h_t)$ is acquired as:
\begin{align}
E(h_t)&\leq a_0+|b_0-a_0|+a_1E(h_{1,t-1})+b_1E(h_{2,t-1})+ \nonumber \\
&(|b_2-b_1+\pi _1-a_2|+a_2)E(h_{t-1})+\sum_{i=0}^{\infty}(\pi_{i+2}-b_2\pi_{i+1})E(h_{t-2-i}) \label{12}
\end{align}
Let $\rho(.)$ denotes the spectral radius of a matrix, then we make the following theorem for the stability condition of the ST-HYGARCH model. 
\begin{Theorem}
Let time series $\{y_t\}$ follows the ST-HYGARCH model defined in relations (\ref{4})--(\ref{8}), then the process is asymptotically stable in variance and $lim_{t\rightarrow \infty}E(y_t^2)\leq \infty$  if $\rho(B)<1$.
\end{Theorem}
\textit{\textbf{Proof:}}
Let's define the following matrices
\begin{equation*}
\mathbf{H_t}=
\begin{bmatrix}
E(h_t)\\
E(h_{1,t})\\
E(h_{2,t})\\
E(h_{t-1})
\end{bmatrix} \qquad
\mathbf{A}=
\begin{bmatrix}
\tau\\
a_0\\
b_0\\
0
\end{bmatrix} 
\end{equation*}
where $\tau=a_0+|b_0-a_0|$ and 
\begin{equation*}
\mathbf{C=}
\begin{bmatrix}
(|b_2-b_1+\pi _1-a_2|+a_2) & a_1 & b_1 &\sum_{i=0}^{\infty}(\pi_{i+2}-b_2\pi_{i+1})B^i\\
0 & a_1+a_2 & 0 & 0 &\\
(b_2-b_1+\pi_1) & 0 & b_1 & \sum_{i=0}^{\infty}(\pi_{i+2}-b_2\pi_{i+1})B^i\\
1 & 0 & 0 & 0 
\end{bmatrix}
\end{equation*}
 By using (\ref{12}) and matrices $H_t$, $A$ and $C$ the following recursive inequality is attained:
 \begin{equation}
 H_t\leq A+C H_{t-1}, \qquad t\geq 0 \label{13}
\end{equation}   
with some initial conditions $H_{-1}$. Iterating inequality (\ref{13}), we get 
\begin{equation}
H_t\leq A\sum_{i=0}^{t-1}C^i+C^tH_0:=D_t
\end{equation}
according to matrix convergence theorem (Lancaster and Tismenetsky \cite{lancaster}) the necessary and sufficient condition for the convergence of $D_t$ when $t\rightarrow \infty$ is $\rho(C)<1$. Under this condition, $C^t\rightarrow 0$ as $t\rightarrow\infty$  and if $(I-C)$ exist then $\sum_{t=0}^{t-1}C^i\rightarrow (I-C)^{-1}$. So if $\rho(C)<1$ 
$$\lim_{t\rightarrow\infty} C_t<(I-C)^{-1}A.$$
 
\section{Estimation}
 Let $\theta=(a_0,a_1,a_2,b_0,b_1,b_2,d,\gamma)^\prime$ denote the parameter vector of the ST-HYGARCH model defined in relations (\ref{4}) - (\ref{8}) and $h_t(\theta)$ refers to the conditional variance of the $y_t$ when the true parameters in ST-HYGARCH model are replaced by the corresponding unknown parameters. Suppose the $y_1,...,y_T$ are a sample from the ST-HYGARCH model defined in (\ref{4}) - (\ref{8}). By assuming the normality on $\epsilon_t$, the conditional log likelihood function is $L(\theta)=-.5 \sum_{t=1}^T l_t(\theta)$ where 
 \begin{align*}
 l_t(\theta)=\ln2\pi+\ln h_t(\theta)+\frac{y_t^2}{h_t(\theta)},
 \end{align*}
 Note that the $h_t(\theta)$ depends on  infinite past observations.  However there are only $T$ observations available in real applications. Hence some initial value are needed, and we may simply assume that $y_s^2=\frac{\sum_{t=1}^T y_t^2}{T}$ for $s\leq0$ (Li et al.\cite{li 2011}).

We employ the quasi-Newton method to find out the maximum likelihood estimator (MLE) of the $\theta$. The derivatives of $L(\theta)$ with respect to the parameters are given as follows:
\begin{align*} \dfrac{\partial L(\theta)}{\partial \theta_{(i)}}=\sum_{t=1}^T\dfrac{1}{2h_t(\theta)}(\dfrac{y_t^2}{h_t(\theta)}-1)\dfrac{\partial h_t(\theta)}{\partial \theta_{(i)}}  
\end{align*}
where $\theta_{(i)}$ refers to the $i-th$ element of the $\theta$. The partial derivatives of $h_t(\theta)$ are obtained as:
\begin{center}
\begin{align*}
&\dfrac{\partial h_t(\theta)}{\partial a_0}=(1-w_t)(1+a_1\dfrac{\partial h_{1t-1}}{\partial{a_0}})\\
&\dfrac{\partial h_t(\theta)}{\partial a_1}=(1-w_t)(h_{1t-1}+a_1\dfrac{\partial h_{1t-1}}{\partial{a_1}})\\
&\dfrac{\partial h_t(\theta)}{\partial a_0}=(1-w_t)(y_{t-1}^2+a_1\dfrac{\partial h_{1t-1}}{\partial{a_2}})\\
&\dfrac{\partial h_t(\theta)}{\partial b_0}=w_t(1+b_1\dfrac{\partial h_{2t-1}}{\partial{b_0}})\\
&\dfrac{\partial h_t(\theta)}{\partial b_1}=w_t(h_{2t-1}+b_1\dfrac{\partial h_{2t-1}}{\partial{b_1}}-y_{t-1}^2)\\
&\dfrac{\partial h_t(\theta)}{\partial b_2}=w_t(b_1\dfrac{\partial h_{2t-1}}{\partial{b_2}}+(1-B)^dy_{t-1}^2)\\
&\dfrac{\partial h_t(\theta)}{\partial d}=w_t\Big(b_1\dfrac{\partial h_{2t-1}}{\partial{d}}-(1-b_2B)(1-B)^d \log(1-B) y_{t}^2 \Big)\\
&\dfrac{\partial h_t(\theta)}{\partial \gamma}=\dfrac{\partial w_t}{\partial \gamma}\Big(h_{t2}-h_{t1}\Big)
\end{align*}
\end{center}  

where $ \dfrac{\partial w_t}{\partial \gamma}=\dfrac{-z_texp(-\gamma z_t)}{(1+exp(-\gamma z_t))^2}$.

\section{Testing Smooth Transition property}
For testing the presence of the smooth transition property in time series we consider the  score test. This test very convenient because  that it dose not require the estimation of the model under alternative hypothesis. It only require  the constrained estimator under $H_0$. The null hypothesis of testing smooth transition property corresponds to testing $H_0: \gamma=0$ against $H_1:\gamma>0$ in the ST-HYGARCH model defined by relations (\ref{4}) - (\ref{8}). Under null hypothesis $w_t=\dfrac{1}{2}$. The null hypothesis implies the absence of the smooth transition property and we obtain standard  HYGARCH model (Amado and Ter\"asvirta \cite{amado and terasvirta  2008}). 
Suppose that $\eta=(a_0,a_1,a_2,b_0,b_1,b_2,d)^\prime$, then $\theta=(\eta^\prime,\gamma)^\prime$. The conditional log-likelihood function can be written as $L(\eta,\gamma)=-0.5\sum_{t=1}^Tl_t(\eta,\gamma)$ when $l_t(\eta,\gamma)=\ln2\pi+\ln h_t(\eta,\gamma)+\frac{y_t^2}{h_t(\eta,\gamma)}$. At bellow the $\sim$ indicates the maximum likelihood estimator under $H_0$.                  

\begin{Theorem}
Suppose that the time series $\{y_t\}$ follow the ST-HYGARCH model defined by relations  (\ref{4}) - (\ref{8}) and  assume that  $\tilde{\theta}= (\tilde{\eta}^\prime,0)^\prime$  is asymptotically normal. Under $H_0:\gamma=0$, the score  test statistic 
\begin{equation}
\psi_{s}=\dfrac{S^2(\tilde{\eta})}{\tilde{\kappa}(Q-R^\prime J^{-1}R)} \label{14}
\end{equation}  
is asymptotically follow the chi-squared distribution with 1 degree of freedom under some regularity conditions. Where $S(\tilde{\eta})=\dfrac{1}{\sqrt{T}}\sum_{t=1}^T\dfrac{\partial l_t(\tilde{\eta},0) }{\partial \gamma}$, $\tilde{\kappa}=\dfrac{1}{T}\sum_{t=1}^T\Big(\frac{y_t^2}{h_t(\tilde{\eta},0)}-1\Big)^2$
$$Q=\frac{1}{T}\sum_{t=1}^T\frac{1}{h_t^2(\tilde{\eta},0)} \Big(\frac{\partial h_t(\tilde{\eta},0)}{\partial\gamma}\Big)^2,$$
$$R=\frac{1}{T}\sum_{t=1}^T\frac{1}{h_t^2(\tilde{\eta},0)} \Big(\frac{\partial h_t(\tilde{\eta},0)}{\partial\gamma}\Big) \Big(\frac{\partial h_t(\tilde{\eta},0)}{\partial\eta}\Big)$$
and
$$J=\frac{1}{T}\sum_{t=1}^T\frac{1}{h_t^2(\tilde{\eta},0)} \Big(\frac{\partial h_t(\tilde{\eta},0)}{\partial\eta}\Big) \Big(\frac{\partial h_t(\tilde{\eta},0)}{\partial\eta^\prime}\Big).$$
\end{Theorem}
\textit{\textbf{Proof:}}
Suppose $\xi_T(\theta)=\dfrac{1}{\sqrt{T}}\sum_{t=1}^T\dfrac{\partial l_t(\theta)}{\partial\theta}$ is the average score test vector and $I(\theta)$ is the population information matrix. Let $H_0:\gamma=0$ and true parameter vector under $H_0$ be $\theta_0=~(\eta_0^\prime,0)^\prime$. The  LM statistic test is defined as follows:
 \begin{equation}
 \psi_{LM}=\xi_T(\tilde{\theta})^\prime I^{-1}(\theta_0)\xi_T(\tilde{\theta})\sim\chi^2_{(1)} \label{15}.
 \end{equation}
Let $\xi_T(\theta)=(\xi_{1T}(\eta^\prime),\xi_{2T}(\gamma))^\prime $ where 

\vspace{.5cm}$\xi_{1T}(\eta)=\dfrac{1}{\sqrt{T}}\sum_{t=1}^T\dfrac{\partial l_t(\eta,\gamma)}{\partial\eta}$ \qquad and \qquad
 $\xi_{2T}(\gamma)=\dfrac{1}{\sqrt{T}}\sum_{t=1}^T\dfrac{\partial l_t(\eta,\gamma)}{\partial\gamma}$.

\vspace{.5cm} Hence
\begin{equation}
\xi_{T}(\tilde{\theta})=(0,\xi_{2T}(0))^\prime, \qquad \xi_{2T}(0)=\dfrac{1}{\sqrt{T}}\sum_{t=1}^T\dfrac{\partial l_t(\tilde{\eta},0) }{\partial \gamma}=S(\tilde{\eta})\label{16}
\end{equation}

 and
\begin{align*}
\dfrac{\partial l_t(\tilde{\eta},0) }{\partial \gamma}=\sum_{t=1}^T(1-\frac{y_t^2}{h_t(\tilde{\eta},0)})\dfrac{1}{h_t(\tilde{\eta},0)}\dfrac{\partial h_t(\tilde{\eta},0)}{\partial\gamma}.
\end{align*}

Under normality, the population information matrix equals to negative expected value of the average Hessian matrix:
\begin{align*}
I(\theta) =E\left[ \dfrac{\partial^2 \log  f(y_t|\Upsilon_{t-1},\theta)} {\partial \theta \partial \theta^\prime} \right]=-E\left[ \dfrac{1}{T}\sum_{t=1}^T\dfrac{\partial^2 l_t(\theta)}{\partial \theta \partial \theta^\prime}\right]=E\left[ \dfrac{1}{T}\sum_{t=1}^T\dfrac{\partial l_t(\theta)}{\partial \theta}\dfrac{\partial l_t(\theta)}{\partial \theta^\prime}\right].   
\end{align*}

Note that  since (\ref{15}) depend on the unknown parameter value $\theta_0$ so it is useless. It is usual to evaluate the $I^{-1}(\theta_0)$ at the $\tilde{\theta}$ to get a usable statistic. 
Hence 
\begin{align}
I(\tilde{\theta})=
\begin{bmatrix}
\tilde{I}_{11}& \tilde{I}_{12}\\
\tilde{I}_{21}& \tilde{I}_{22}  \label{17}
\end{bmatrix}
\end{align}
where
 \begin{align}
\tilde{I}_{11}=\tilde{\kappa}J,\qquad \tilde{I}_{12}=\tilde{I}_{21}=\tilde{\kappa}R, \qquad  \tilde{I}_{22}=\tilde{\kappa}Q. \label{18}
 \end{align}
By substituting the (\ref{16}) - (\ref{18}) in (\ref{15}) we get the (\ref{14}).

\section{\textbf{Simulation Study}}
In this section we conduct two simulation experiments to investigate the  performance of the MLE in section 4 and the score test in section 5. Three sample lengths n=500, 1000 and 2000 observations have been used in 
two experiments, and there are 1000 replications for each sample size. In each generated sequence the first 1000 observations have been discarded to avoid the initialization effects, so there are 1000+n observations generated each time.

In the first experiment the data are generated from ST-HYGARCH model defined in (\ref{4})-~(\ref{8}) where $\{\epsilon_t \}$ are \textit{iid} standard normal variables and the value  of the parameter vector are 
$$\theta=(a_0,a_1,a_2,b_0,b_1,b_2,d,\gamma)^\prime=(.35,.30,.40,.10,.20,0,.60,1.50)^\prime.$$
The maximum likelihood estimations  are calculated, the Bias and the root mean squared error (RMSE) are summarized in table 1. It can be seen that both Bias and RMSE are generally small and  decrease as the  sample size increases.
In the second experiment the size and power of the proposed score test in section 5 are investigated. The data generated from ST-HYGARCH model (\ref{4}) - (\ref{8}) when 
$$\theta=(a_0,a_1,a_2,b_0,b_1,b_2,d,\gamma)^\prime=(.35,.30,.40,.10,.20,0,.60,\gamma)^\prime.$$
 $\gamma=0$ corresponds to the size and  $\gamma>0$ correspond to the power of the test. we consider three different value $\gamma$=0.4, 2 and 7 and two significance values .05 and .10. The empirical rejection rates are reported in  table 2. It can be observed that the empirical sizes are all close to the nominal levels and this closeness increases as the sample size increases also empirical powers are increasing function of the sample size and of the $\gamma$.

\begin{table}[t] 
\caption{Estimation results of the ST-HYGARCH model based on 1000 replications. \label{Diff_c}}
\begin{small}
\begin{center}
\begin{tabular}{c c c c c c c c c c c c c }
\hline \hline
   & & & \multicolumn{2}{c}{n=500} & & \multicolumn{2}{c}{n=1000} & & \multicolumn{3}{c}{n=2000} \\
 \cline{3-5} \cline{7-9} \cline{11-13}
 parameter & Real value  & & Bias&  RMSE & &  Bias& RMSE & &&  Bias& RMSE &\\
\hline
$a_0$&0.35 &  & 0.0260 & 0.0846 & & 0.0260 & 0.0715 & && 0.0100 & 0.0274 \\
$a_1$&0.30 &  & 0.0357 & 0.0480 & & 0.0350 & 0.0417 & && 0.0350 & 0.0407 \\
$a_2$&0.40 &  & 0.0079 & 0.0410 & & 0.0078 & 0.0339 & && 0.0065 & 0.0284 \\
$b_0$&0.10 &  & 0.0464 & 0.0514 & & 0.0463 & 0.0495 & && 0.0448 & 0.0497 \\
$b_1$&0.20 &  & -0.0172 & 0.0352 & & -0.0166 & 0.0309 & && -0.0117 & 0.0232 \\
$d$  &0.60 &  & 0.0064 & 0.0407 & & 0.0050 & 0.0436 & && 0.0007 & 0.0263 \\
$\gamma$& 1.50&&0.0227 & 0.0898 & & 0.0179 & 0.0702 & &&0.0030  & 0.0603  \\
 \hline
\end{tabular}
\end{center}
\end{small}
\end{table}

\begin{table}[t] 
\caption{Empirical rejection rates of the score test for the ST-HYGARCH model based on 1000 replications for two significance level 0.05 and 0.10.  $\gamma=0$ corresponds to the size and  $\gamma>0$ correspond to the power of the test. }
\begin{small}
\begin{center}
\begin{tabular}{c c c c c c c c c c c }
\hline \hline
   & &  \multicolumn{2}{c}{n=500} & & \multicolumn{2}{c}{n=1000} & & \multicolumn{2}{c}{n=2000} \\
 \cline{3-4} \cline{6-7} \cline{9-10}
 $\gamma$   & & 0.05&  0.10 & &  0.05& 0.10 & &  0.05& 0.10 &\\
\hline
$0$ &  & 0.058 & 0.128 & & 0.051 & 0.111 & & 0.051 & 0.101 \\
$0.4$ &  & 0.113 & 0.189 & & 0.120 & 0.198 & & 0.144 & 0.252 \\
$2$ &  & 0.180 & 0.279 & & 0.191 & 0.286 & & 0.234 & 0.347 \\
$7$ &  & 0.212 & 0.310 & & 0.235 & 0.328 & & 0.288 & 0.382 \\
 \hline
\end{tabular}
\end{center}
\end{small}
\end{table}

\section{Real Data}
In this section, we apply the proposed ST-HYGARCH model as well as HYGARCH model to the daily log returns (in percentage) of the {\textit{S}\&\textit{P}500} indices. There are 1500 observations from February 17, 2009 to January 30, 2015. Figure 1  presents the sample path and the conditional variances of the data, which show evidences of  continues regimes. In table 3   the descriptive statistics of the data are reported. We observe the means are close to zero and also a slightly negative skewness and the common excess kurtosis of the data. We consider three different STHYGARCH models, STHYGARCH(1), STHYGARCH(2) and STHYGARCH(3) respectively corresponding to three different transition variables, $z_{t(1)}=y_{t-1}$, $z_{t(2)}=h_{t-1}$  and
 \begin{equation*}
 z_{t(3)}=\Big\lbrace
 \begin{array}{lr}
 y_{t-1} & \textit{if}\qquad y_{t-1}<p(95)\\
 \dfrac{y_{t-1}+y_{t-2}+y_{t-3}}{3} & \textit{if}\qquad y_{t-1}>p(95) \\
\end{array} 
\end{equation*} 
where $p(95)$ refers to the 95-th percentile of the squared returns. In $z_{t(3)}$ the asymmetry effect of the size is more stressed. Firstly, we applied the proposed score test $\psi_s$ to data. The results are reported in table 4. It can be observed that the hypothesis $H_0:\gamma=0$ is rejected for all models at 5\% significance level ($\chi^2_{(0.05,1)}=3.86$). Secondly, we compare the ability of different models in computing true conditional variances which are measured by squared observations. We have used the first 1000 observations as in-sample data to estimate the models, and the remaining 500 observations as out-of sample data to perform forecasting. Table 5  provides the maximum likelihood estimates of the models. To evaluate the performance of the different models, we calculated the RMSE and the Log Liklihood value (LLV). Results are given in the table 6. As out-of-sample performance the one-day-ahead forecasts are computed using estimated models. From table 6 it can be seen that the ST-HYGARCH models out perform the HYGARCH model and also the ST-HYGARCH(3) model has the lower RMSE and  higher LLV than to other models. So it seems that using the transition variable $z_{t(3)}$ the ST-HYGARCH model can move between different regimes as well. Note that in ST-HYGARCH(2) the signs of the observations are ignored and this led to  the weaker results. So it  mean that the size and sign of  past observations have noticeable influence in the improvement  the smooth transition structure. To clarifying the out-performances of the different ST-HYGARCH models we plot the the forecasting conditional variances and true conditional variances (squared returns) for some of data in figure 2,  it shows that ST-HYGARCH models better forecast the true conditional variances than HYGARCH. Figure 3 displays the absolute forecasting error of the different models for some of data. It can be observed the ST-HYGARCH(3) model have lower errors than to other models.

\begin{table}[t]
\caption{Descriptive statistics of  \textit{S}\&\textit{P}500 daily  returns}
\begin{small}
\begin{center}
\begin{tabular}[t]{c c c c c c c} \hline
series&Mean&Std.devd&Minimum&Maximum&Skewness&Kurtosis \\ \hline
\textit{S}\&\textit{P}&0.062&1.114&-6.896&6.837&-0.148&4.564 \\
 \hline
\end{tabular}
\end{center}
\end{small}
\end{table}

\begin{table}[t]
\caption{Score test statistic value for  {\textit{S}\&\textit{P}500} daily returns }
\begin{small}
\begin{center}
\begin{tabular}[t]{c c c c c c } \hline
 &ST-HYGARCH(1)&ST-HYGARCH(2)& ST-HYGARCH(3)\\ \hline 
$\psi_s$&5.535&4.362&5.601\\ \hline
\end{tabular}
\end{center}
\end{small}
\end{table}

\begin{table}[!hpbt]
\caption{Maximum likelihood estimates of ST-HYGARCH(1), ST-HYGARCH(2), ST-HYGARCH(3) and  HYGARCH models on {\textit{S}\&\textit{P}500}daily returns. }
\begin{small}
\begin{center}
\begin{tabular}{c c c c c c c  c c c}
\hline
 & ST-HYGARCH(1) &ST-HYGARCH(2)& ST-GARCH(3)&HYGARCH  \\
\hline
$a_{0}$&0.307 &0.366& 0.222&0.424\\
$a_{1}$  &0.233&0.218& 0.249&0.253\\
$a_{2}$ &0.492&0.486&0.459&0.435\\
$b_{0}$&0.446&0.433&0.241&0.427\\
 $b_{1}$ & 0.139&0.177&0.124&0.185\\
 $d$ &0.875&0.876&0.930&0.577\\
 $w$&0.393&0.620&0.254&0.285\\
 \hline
\end{tabular}
\end{center}
\end{small}
\end{table}

\begin{table}[t] 
\caption{Measures of performance  of ST-HYGARCH(1), ST-HYGARCH(2), ST-HYGARCH(3) and  HYGARCH models on {\textit{S}\&\textit{P}500}daily returns}
\begin{small}
\begin{center}
\begin{tabular}{c c c c c c c c c  }
\hline \hline
   & &  \multicolumn{2}{c}{In-Sample} & & \multicolumn{2}{c}{Out-of-Sample}  \\
 \cline{3-4} \cline{6-7} 
 Model   & & RMSE&  LLV & &  RMSE& LLV \\
\hline
ST-HYGARCH(1) &  & 1.312 & -1218.6 & & 0.494 & -493.4 \\
ST-HYGARCH(2) &  & 1.564& -1219.5 & & 0.530 & -499.0  \\
ST-HYGARCH(3) &  & $1.258^*$& $-1135.9^*$ & & $0.367^*$ & $-439.1^*$ \\
HYGARCH         &    &  1.864& -1271.3& & 0.641 & -520.1\\
 \hline
\end{tabular}
\end{center}
\end{small}
\end{table}

\begin{center}
\begin{figure}[!btp]
\includegraphics[width=17cm]{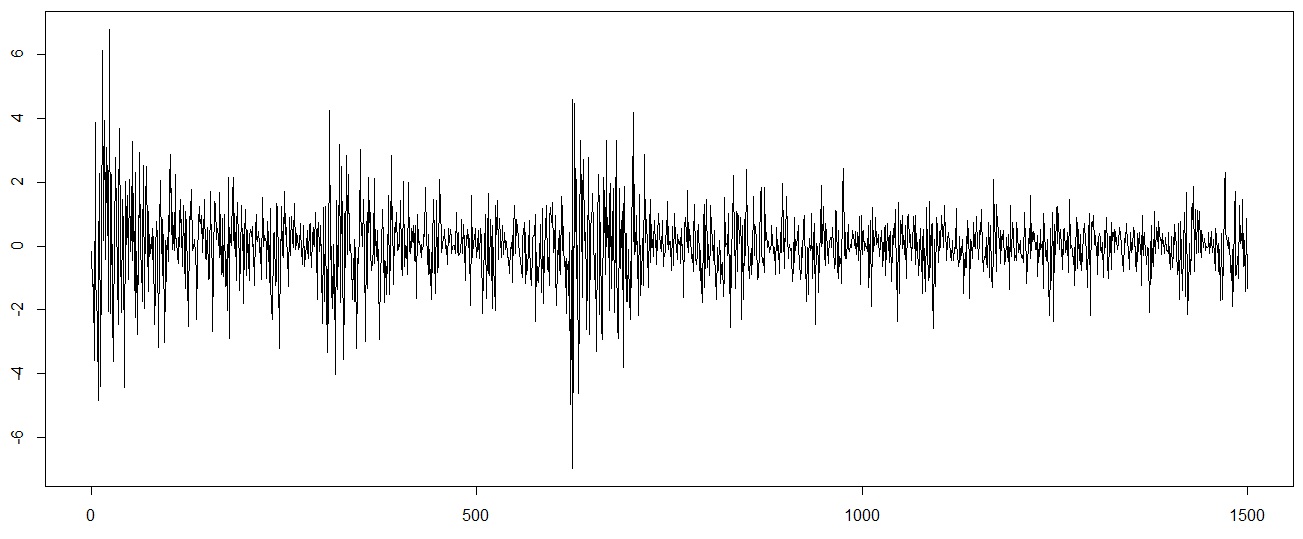}
\includegraphics[width=17cm]{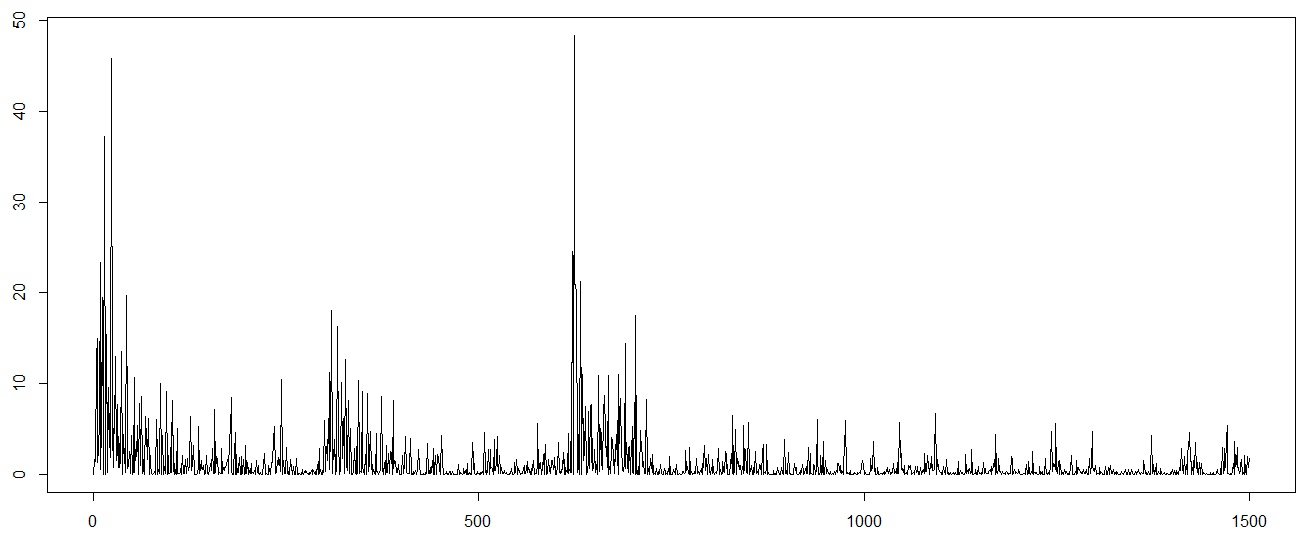}

\caption{(Up): Log returns of \textit{S}\&\textit{P}500 daily data. (Bottom):Conditional variances (squared returns ) of \textit{S}\&\textit{P}500 daily data.} 
\end{figure}
\end{center}

\begin{center}
\begin{figure}[!btp]
\includegraphics[width=17cm]{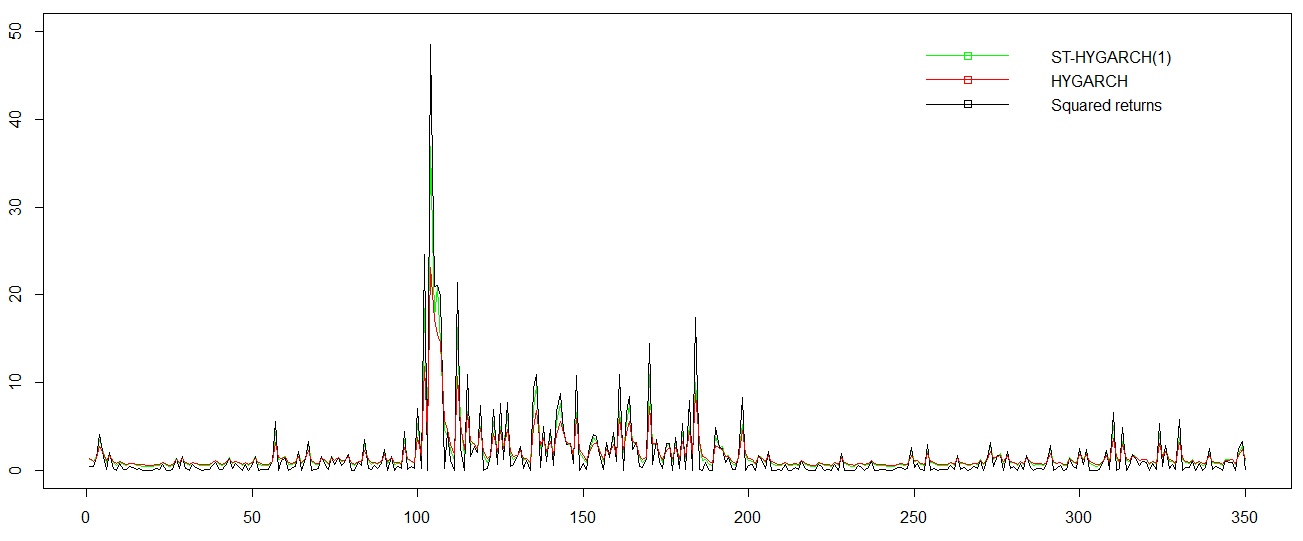}
\includegraphics[width=17cm]{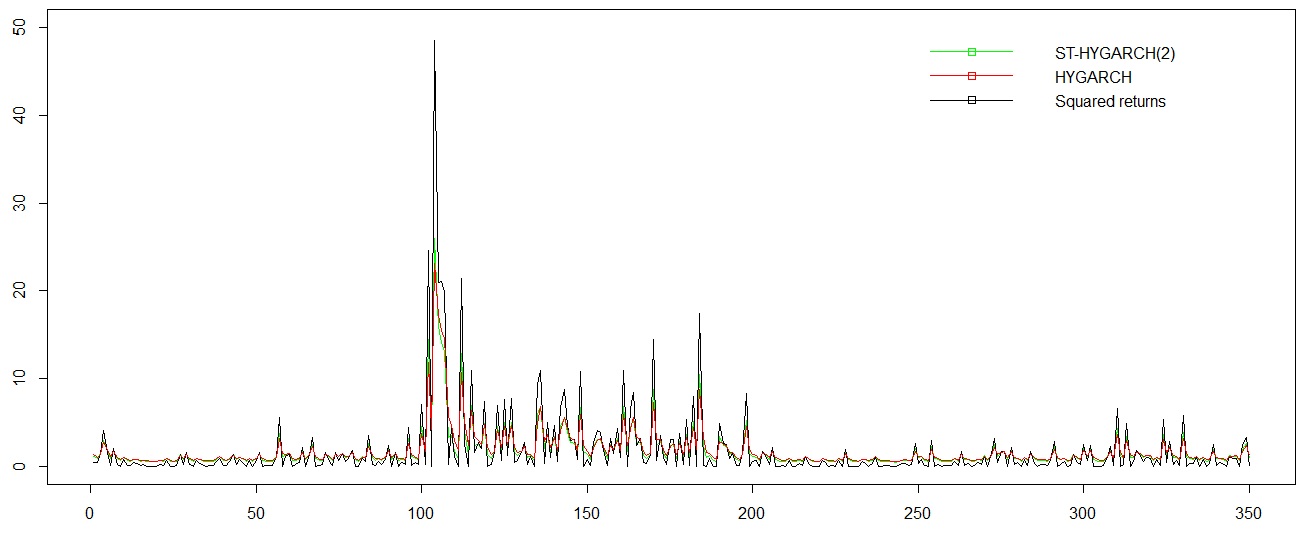}
\includegraphics[width=17cm]{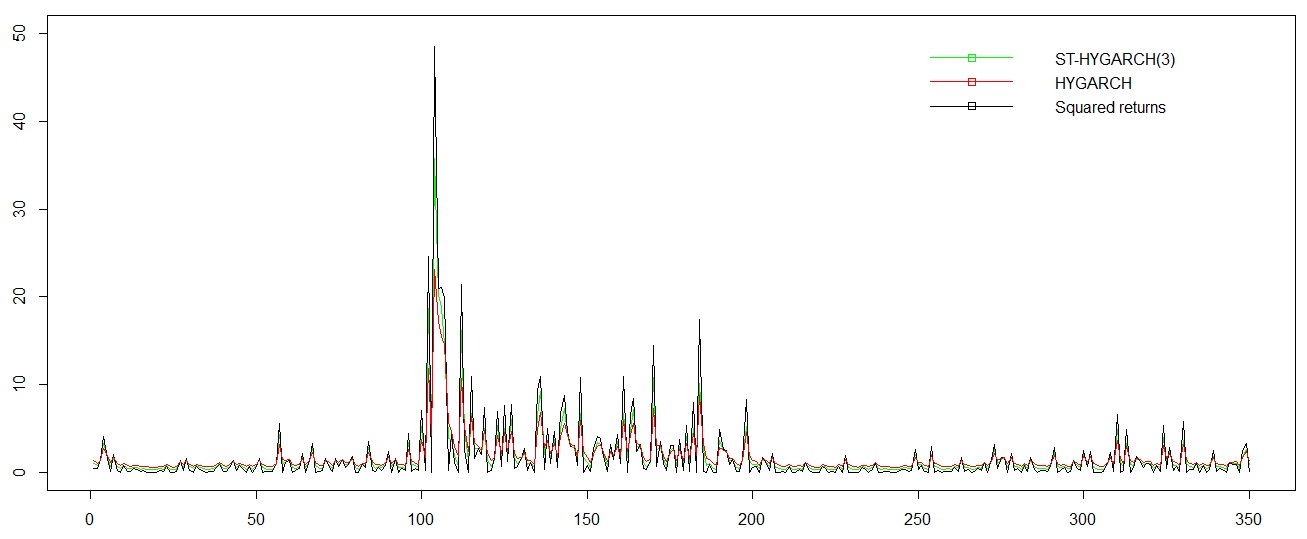}
\caption{:Squared returns and forecasting conditional variances with ST-HYGARCH models and HYGARCH  model for  some of \textit{S}\&\textit{P}500  daily returns.} 
\end{figure}
\end{center}

\begin{center}
\begin{figure}[!btp]
\includegraphics[width=17cm]{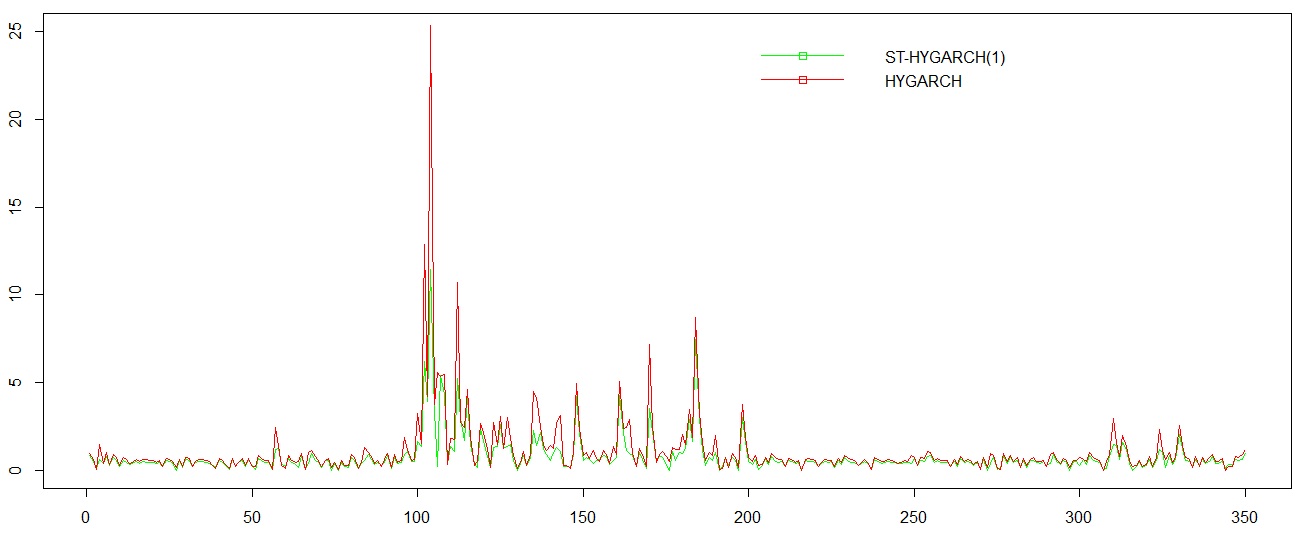}
\includegraphics[width=17cm]{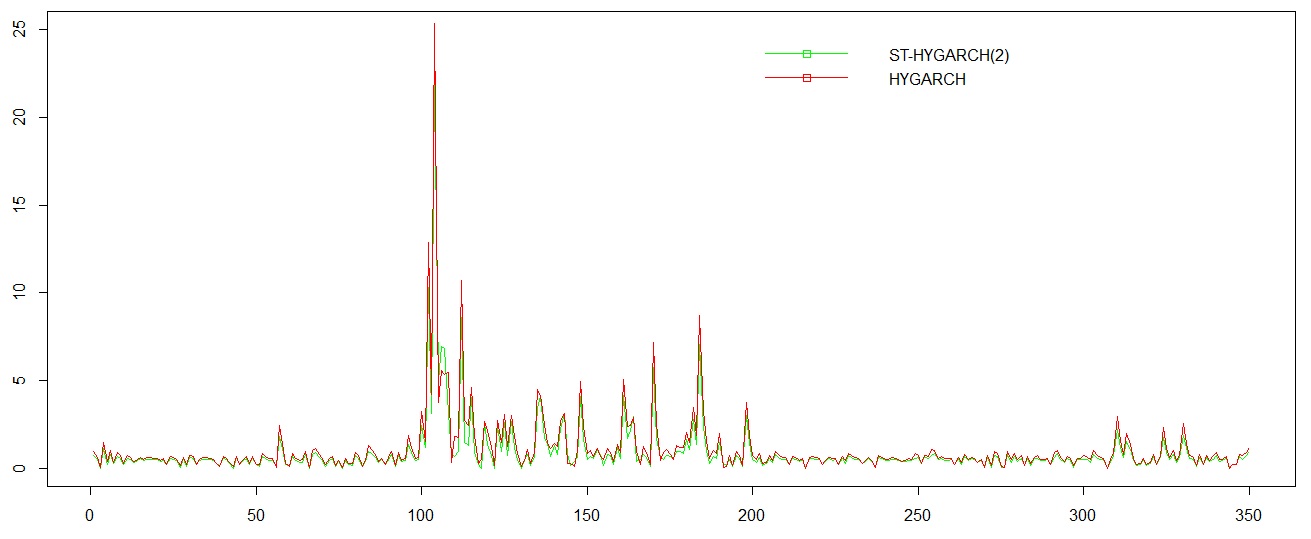}
\includegraphics[width=17cm]{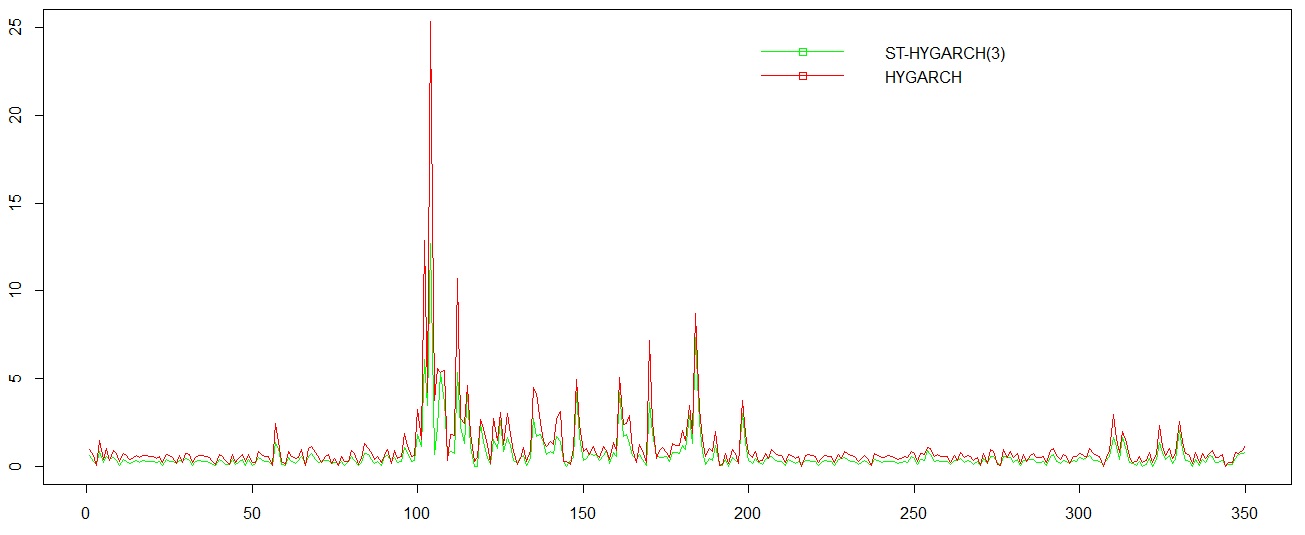}
\caption{Absolute forecasting errors between squared returns and forecasting conditional variances with ST-HYGARCH models and HYGARCH  model for  some of \textit{S}\&\textit{P}500  daily  returns}.
\end{figure}
\end{center}

\newpage
\noindent
\section{Conclusion}

In this paper we study an extension on  HYGARCH model, say ST-HYGARCH which has smooth time-varying structure. This model is capable to capture  different  volatility levels using logistic function as a transition tool. ST-HYGARCH model is flexible to capture long and short memory  volatilities. Such behavior often occurs  in many  financial time series.  Model is more realistic by its time-varying  structure. We showed the ST-HYGARCH model is asymptotically stable. One of the privilege of this work is implying of score test to check existence of such Smooth transition structure. Simulation evidences showed that empirical performance of test is competitive. Application of score test  to the \textit{S}\&\textit{P}500 indices rejects  HYGARCH in favour of  ST-HYGARCH one. Applying on  \textit{S}\&\textit{P}500 data, we find that  ST-HYGARCH models out-perform the HYGARCH model in forecasting.  Asymmetric behavior and heavy tailed property of financial time series can motivate further researches.


\end{document}